\documentclass[aip,jcp,reprint]{revtex4-1} 
\usepackage{graphicx}

\usepackage{amsmath}    
\usepackage{graphicx}   
\usepackage{color}
\usepackage{ulem}
\usepackage{caption}
\usepackage{subcaption}

\usepackage{array}
\usepackage{booktabs}
\usepackage{multirow}

\usepackage{booktabs}

\begin{document}
	\title{ A perturbative solution to metadynamics ordinary differential equation}
	\author{Pratyush Tiwary}
	 \email{pt2399@columbia.edu}   
	 \affiliation{Department of Chemistry, Columbia University, New York 10027, USA.}
	
	\author{James F. Dama}
	\affiliation{Department of Chemistry, The James Franck Institute, Institute for Biophysical Dynamics, and Computation Institute, The University of Chicago, Chicago, Illinois 60637, USA.}
	
	\author{Michele Parrinello}
	\affiliation{Department of Chemistry and Applied Biosciences, ETH Zurich, and Facolt\`{a} di Informatica, Instituto di Scienze Computationali,
	Universit\`{a} della Svizzera italiana (USI), Via Giuseppe Buffi 13, CH-6900,
	Lugano, Switzerland. }

	\date{\today}

	\begin{abstract}
	
	Metadynamics is a popular enhanced sampling scheme wherein by periodic application of a repulsive bias, one can surmount high free energy barriers and explore complex landscapes. Recently metadynamics was shown to be mathematically well founded, in the sense that the biasing procedure is guaranteed to converge to the true free energy surface in the long time limit irrespective of the precise choice of biasing parameters. A differential equation governing the post-transient convergence behavior of metadynamics was also derived. In this short communication, we revisit this differential equation, expressing it in a convenient and elegant Riccati-like form. A perturbative solution scheme is then developed for solving this differential equation, which is valid for any generic biasing kernel. The solution clearly demonstrates the robustness of metadynamics to choice of biasing parameters and gives further confidence in the widely used method.
	\end{abstract}
	\maketitle
	\section{Introduction}
	
	Molecular Dynamics (MD) simulations have been very popular over the last few decades for studying the static and dynamic properties of a variety of systems in physical, chemical and biological sciences. However, for many real-world systems MD simulations are plagued by sampling issues \cite{voter_review}. Most such systems can be characterized by deep stable free energy basins separated by high barriers. As such, a Boltzmann-weighted sampling is not sufficient to accurately sample the free energy landscape, as the system mostly stays trapped in a small region of the configuration space. To efficiently sample such landscapes within computer time restraints, over the years several non-Boltzmann sampling paradigms have been proposed \cite{Hansmann2002a,Torrie1977a,fep_rest,Darve2001,Huber1994,valsson2014variational,muller2002predicting,voter1997method,Voter:1997,Tiwary:2013,Grubmuller:1995,sisyphus2,voter_review,laio2002escaping}. 
	
	Metadynamics \cite{laio2002escaping,wtm,tiwary_rewt,meta_ode,meta_review} is one such technique that builds a memory-dependent bias potential and hence continuously changes the sampling weight through the course of the simulation. By gradually enhancing the fluctuations of certain carefully chosen low-dimensional order parameters, commonly known as collective variables, the system is coaxed to visit and spend time in regions of the configuration space that it would ordinarily not visit or visit only rarely. Recently \cite{meta_ode} it was rigorously demonstrated that the discrete protocol of adding bias every certain number of integration steps can be mapped into an ordinary differential equation. It was also shown that this differential equation has a unique fixed point, and as such in the long-time limit metadynamics always gives the true underlying free energy as a function of the deposited bias.
	
	In this short communication, our aim is three-fold as we revisit the differential equation of Ref.  \onlinecite{meta_ode}. First, we simplify it and express it in a Riccati-like form \cite{riccati}. Second, we develop a perturbative scheme that allows us to look not just at the eventual fixed points of this differential equation, but actually obtain the dominant solution at any simulation time and for any biasing form, as long as the kernels are reasonably narrow for a perturbation scheme to apply. Third, from our perturbative solution we demonstrate the robustness of metadynamics free energy estimators  \cite{wtm,tiwary_rewt} that have been in use over the last decade, with respect to precise choice of the biasing form and parameters, and even when far from the infinite time limit needed in Ref.  \onlinecite{meta_ode}. Note that the results of this work are not an attempt to calculate the rate at which metadynamics converges. Instead, this communication shows that for all practical purposes, the free energy estimator in popular use for metadynamics that was derived assuming delta-like kernels, is actually the dominant part of the true free energy estimator derived here for generic kernels of narrow but finite width. The results are illustrated through a simple 1-d model landscape. The mathematics described in this article should further re-enforce the confidence in using metadynamics for sampling complex free energy landscapes.

	\section{Background: the metadynamics differential equation}
	
	Let $ s(\textbf{R}) $ denote a low-dimensional collective variable (CV) that is a function of the system coordinates $\textbf{R}$ at a temperature $T$ of interest. Metadynamics \cite{meta_laio,wtm} involves building a history-dependent bias $V_{n}(s)$ in the CV space where $n$ denotes the number of MD integration steps that have been carried out, and $ s_n $ denotes the value of $ s(\textbf{R})$ at iteration $n$. The precise rule for the update of the bias is 
	\begin{equation}
	V_{n+1}(s) = V_n (s) + G(s,s_{n+1})e^{-{V_n (s_{n+1}) \over \Delta T}}
	\label{eq:update_net}
	\end{equation}
	
	Here $G(s,s_{n+1})$ is the hill function or biasing kernel deposited at time $t_{n+1}$ and centered around $s=s_{n+1}$. The points $s_n$ are sampled as per the time-dependent probability density $P(s,t)$ that evolves due to the bias addition. Through the use of a tempering parameter $\Delta T$,  the amplitude of the hill function each time a point is re-visited can be tuned down, depending on the bias already deposited \cite {wtm}. Often this tempering factor is expressed in the form of the so-called biasing factor\cite{wtm,tiwary_rewt} $\gamma =  { T+\Delta T \over T  }$ .
	
	 It was recently proven  \cite{meta_ode} that for any generic form of the biasing kernel $G(s,s')$, metadynamics is guaranteed to eventually converge to the final state that it was designed to converge to in its original formulations \cite{laio2002escaping,wtm}. We will first summarize the broad outlines of the technical proof. A complete description has been given in Ref.  \onlinecite{meta_ode} and the interested reader is referred to it for further details.   
	 
	 In their proof, Dama and co-authors divide the total bias at any point $s$ into an average bias $\bar{V_n} = {\int V_n (s) ds \over \int ds} $and a so-called driving bias $\tilde{V_n} (s) = V_n(s) - \bar{V_n}$. Note that the average bias is a kernel-dependent offset \cite{bussi2006} that is same for all $s$ values, and it amounts to lifting up the whole landscape by a constant without affecting the relative probabilities. As such it is irrelevant for our purposes, and it suffices to understand the effect of $\tilde{V_n} (s)$. The update scheme that results for this driving bias is of the form:
	\begin{equation}
	\tilde{V}_{n+1}(s) = \tilde{V}_n (s) + e^{-{\bar{V}_n \over \Delta T}} \Gamma (s,s_{n+1}, \tilde{V}_n)
	\label{eq:update_drive}
	\end{equation}
	
	where $\Gamma$ is a functional of the biasing kernel $G$ and the bias $V_n (s)$, the full expression for which can be found in Ref.  \onlinecite{meta_ode}. Take note of the structure of Eq. \ref{eq:update_drive}: It is of the generic form $\theta_{n+1}= \theta_{n} + \epsilon_n Y_n (\theta_n,x_n)$. Here $ \theta_{n}$  are the iterating values of the driving bias, $\epsilon_n=e^{-{\bar{V}_n \over \Delta T}} $ are step sizes, and $Y_n (\theta_n,x_n)$ are history-dependent updates.
	
	For such a generic stochastic iteration, there is a result described in Ref. \onlinecite{meta_ode} that says that this discrete stochastic iteration converges like the solution of the differential equation ${d\theta \over d \nu} = \langle Y(\theta,x)\rangle $ under certain conditions, where $\nu = \Sigma \epsilon_n$ and the expectation is taken over the equilibrium distribution of $x$ given $\theta$. Ref.  \onlinecite{meta_ode}  describes these conditions in detail and why metadynamics as per Eq. \ref{eq:update_drive} satisfies them - here we mention just one of these. As per this particular condition, the differential equation will hold if the sum of $\epsilon_n$ diverges, the sum of $\epsilon_n^2$ converges and the $Y_n$ are bounded. It can be demonstrated as done in Ref.  \onlinecite{meta_ode} that these conditions hold true for metadynamics given the terms in Eq. \ref{eq:update_drive} and setting $\epsilon_n = e^{-{\bar{V}_n \over \Delta T}} $. It is remarkable that at least in the authors' knowledge, no other timescale definition in Eq. \ref{eq:update_drive} seems to meet these divergence-convergence criteria. Thus, the discrete operation of metadynamics can be mapped in the long time limit to a differential equation which after substituting various terms (Ref.  \onlinecite{meta_ode}) and some further straightforward algebraic manipulations turns out to be:
	\begin{align}
	\label{eq:gen_bussi}
	{d {V}(s,t) \over dt} 
	= \int ds' e^{-{V(s',t) \over \Delta T}} G(s,s') p_b(s',t) 
	\end{align}
	
	Here $p_b(s,t)  = {    e^{-{  F(s) + V(s,t) \over T }}   \over \int ds' e^{-{  F(s') + V(s',t) \over T }}    }   $ is the biased equilibrium distribution and $t$ is the simulation time. 
	
	This equation contains an exponential non-linearity, but we will now show it can be simplified into an equation with only a simpler quadratic non-linearity. We begin by introducing a second timescale $\tau$ defined as
	\begin{align}
	\label{eq:tau}
	\tau = \int_0^t dt' \left[  \int ds e^{-V(s,t') \over T} p(s)   \right] ^{-1} \equiv \int_0^t dt' e^{c(t') \over T}
	\end{align}
	
	where $p(s)$ is the unbiased probability distribution of the system, and we have introduced the function $c(t)$ \cite{tiwary_rewt,bonomi_rewt}, that is a lower bound estimate of the reversible work done on the system through the course of metadynamics:
	\begin{align}
	\label{eq:coft}
	c(t) \equiv T \log \frac{ \int ds e^{-\beta F(s)} }{ \int ds e^{-\beta (F(s) + V(s,t))} }
	\end{align}
	
	Such a lower bound becomes tighter and tighter as the simulation progresses. By expressing Eq. \ref{eq:gen_bussi} in terms of $\tau$ and working in the units of time so that $ \omega \gamma / \Delta T=1$, we then obtain a simpler-looking (but not so easy to solve) integro-differential Riccati-like equation \cite{riccati} that governs metadynamics for any generic biasing kernel $G$:
	\begin{align}
	\label{eq:gen_bussi_2}
	{d {y}(s,\tau) \over d \tau} 
	= -y(s,\tau) \int ds' G(s,s') y(s',\tau)
	\end{align}
	
	where $y(s,\tau) \equiv p(s) e^{-\gamma V(s,\tau) / \Delta T}$. In the rest of this communication, we will be solving Eq. \ref{eq:gen_bussi_2}.
	
	\section{Solution}
	
	We first solve Eq. \ref{eq:gen_bussi_2} for the special case when $G(s,s')=\delta(s,s')$. This can be interpreted as the zero width biasing kernel limit. In this case which was also solved in Ref. \onlinecite{tiwary_rewt}, the convolution in Eq. \ref{eq:gen_bussi_2} integrates out with standard methods \cite{tiwary_rewt} and we obtain as solution:
	\begin{align}
	\label{eq:0order}
	y_0(s, \tau) &= \frac{p(s)}{1 + \tau p(s)}
	\end{align}
	
	where the subscript 0 denotes 0-th order solution. Given that this delta-like kernel limit is tractable, we propose to solve for the general case of Eq. \ref{eq:gen_bussi_2} by considering a perturbation series around the difference $G(s,s')-\delta(s,s')$. This will be a singular perturbation series valid only in the limit of narrow hills, and we do not characterize the radius of convergence in this publication except by example. However, from an intuitive unit analysis it is clear that the terms scale as the length-scale of the added hills divided by the length-scale of the underlying probability distribution, ensuring that this series has a clear physical sense as a narrow-hill limit and thus that the radius of convergence in $G(s,s')$ is in principle meaningful and finite for any $p(s)$ that could be encountered in practice.

	Let us write the true solution to Eq. \ref{eq:gen_bussi_2} as
	\begin{align}
	\label{eq:soln_true}
	y(s,\tau) = \sum\limits_{n=0}^{n=\infty} \text{ }\epsilon^n y_n(s,\tau)
	\end{align}
	where we have introduced a bookkeeping parameter $\epsilon$ for each order in $G-\delta$, that we will later set as 1. By putting Eq. \ref{eq:soln_true} in Eq. \ref{eq:gen_bussi_2} and equating terms with same power in $\epsilon$ we obtain
	\begin{equation} 
	\begin{split}
	{d  y_n(s,\tau)    \over d\tau  } &=  -  \sum\limits_{i=0}^{n} y_i(s,\tau)  y_{n-i}(s,\tau)  - \\
	&    \sum\limits_{i=0}^{n-1} y_i(s,\tau)     \int ds'  \left( G(s,s')-\delta(s,s')  \right) y_{n-i-1}(s',\tau)
	\end{split}
	\end{equation}
	To solve this equation we note that for any $n$ it is of the generic form
	\begin{equation} 
	\label{linear_ode}
	\begin{split}
	{d y_n(s,\tau) \over d\tau} + 2y_0(s,\tau)y_n(s,\tau) = f_n(s,\tau)
	\end{split}
	\end{equation}
	which is a first order linear ordinary differential equation with an inhomogeneous term $f_n$ set by the lower-order terms of the solution as
	\begin{equation} 
	\begin{split}
	f_n(s, \tau) &= -[\sum_{i=1}^{n-1} y_i (s, \tau)  y_{n - i} (s, \tau)] - \\
	 & \sum_{i=0}^{n-1} y_i(s, \tau) \int ds' (G(s,s') - \delta(s, s')) y_{n - 1 - i}(s', \tau)
	\end{split}
	\end{equation}
	By using Eq. \ref{eq:0order}, we can then solve Eq. \ref{linear_ode} to obtain
	\begin{equation} 
	\begin{split}
	\label{gen_sol}
	y_n(s, \tau) &= \frac{1}{(1 + \tau p(s))^2} \int_0^\tau (1 + \tau' p(s))^2 f_n(s,\tau') d\tau'
	\end{split}
	\end{equation}
	So far the perturbative solution is exact and can be solved in an iterative manner by gradually ascending in $n$. We now look at the dominant term in the first order $G-\delta$ correction. That is given by
	\begin{equation} 
	\begin{split}
	y_1(s,\tau) &= \frac{p(s)^2}{(1 + \tau p(s))^2} . \\
	&	\int ds' (G(s, s') - \delta(s, s')) {1\over p(s')} \log[1 + \tau p(s')] 
	\end{split}
	\end{equation}
	
				\begin{figure}
    \begin{subfigure}{0.45\textwidth}
        \includegraphics[height=1.85in]{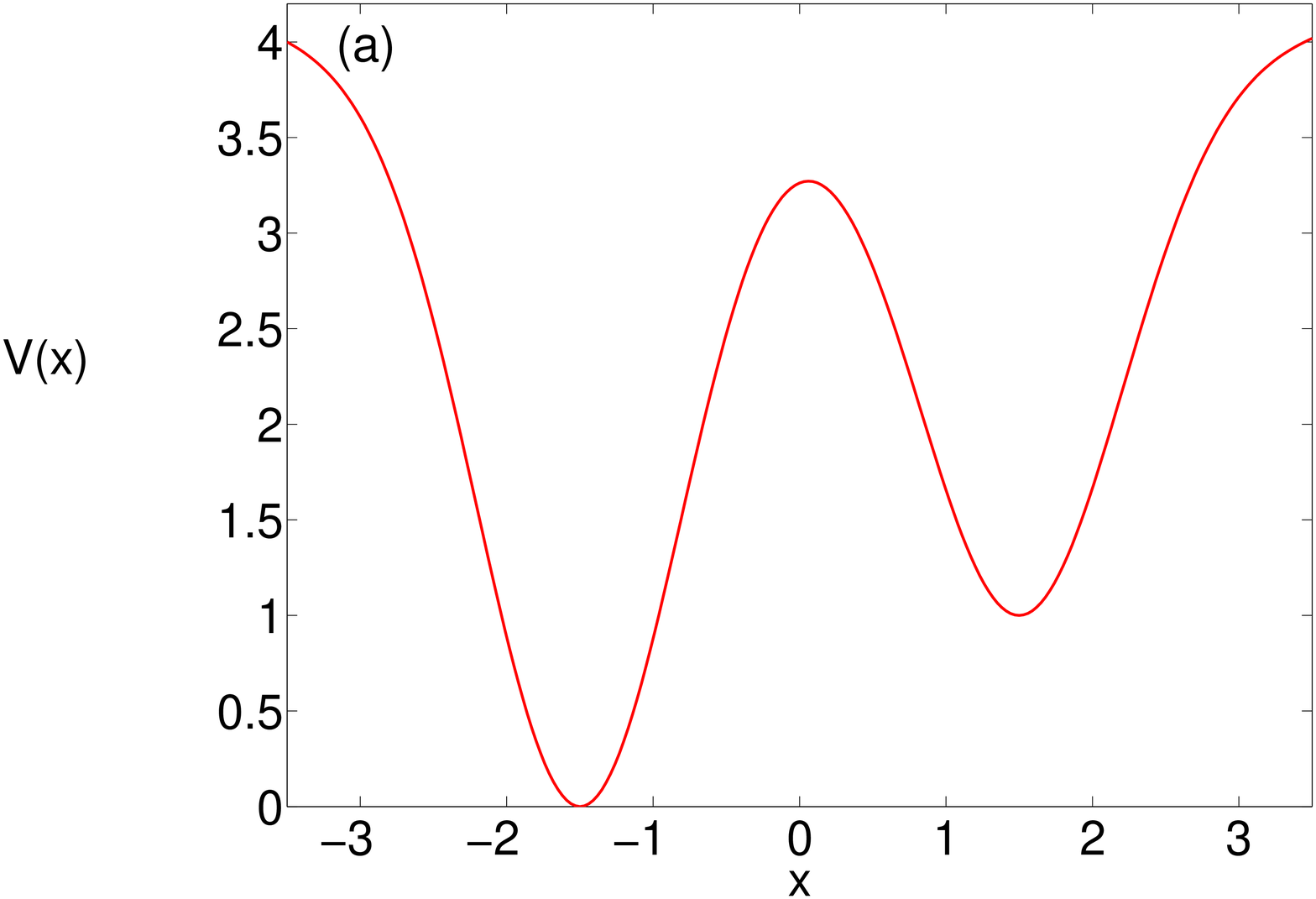}
    \end{subfigure}
    \begin{subfigure}{0.45\textwidth}
        \includegraphics[height=1.85in]{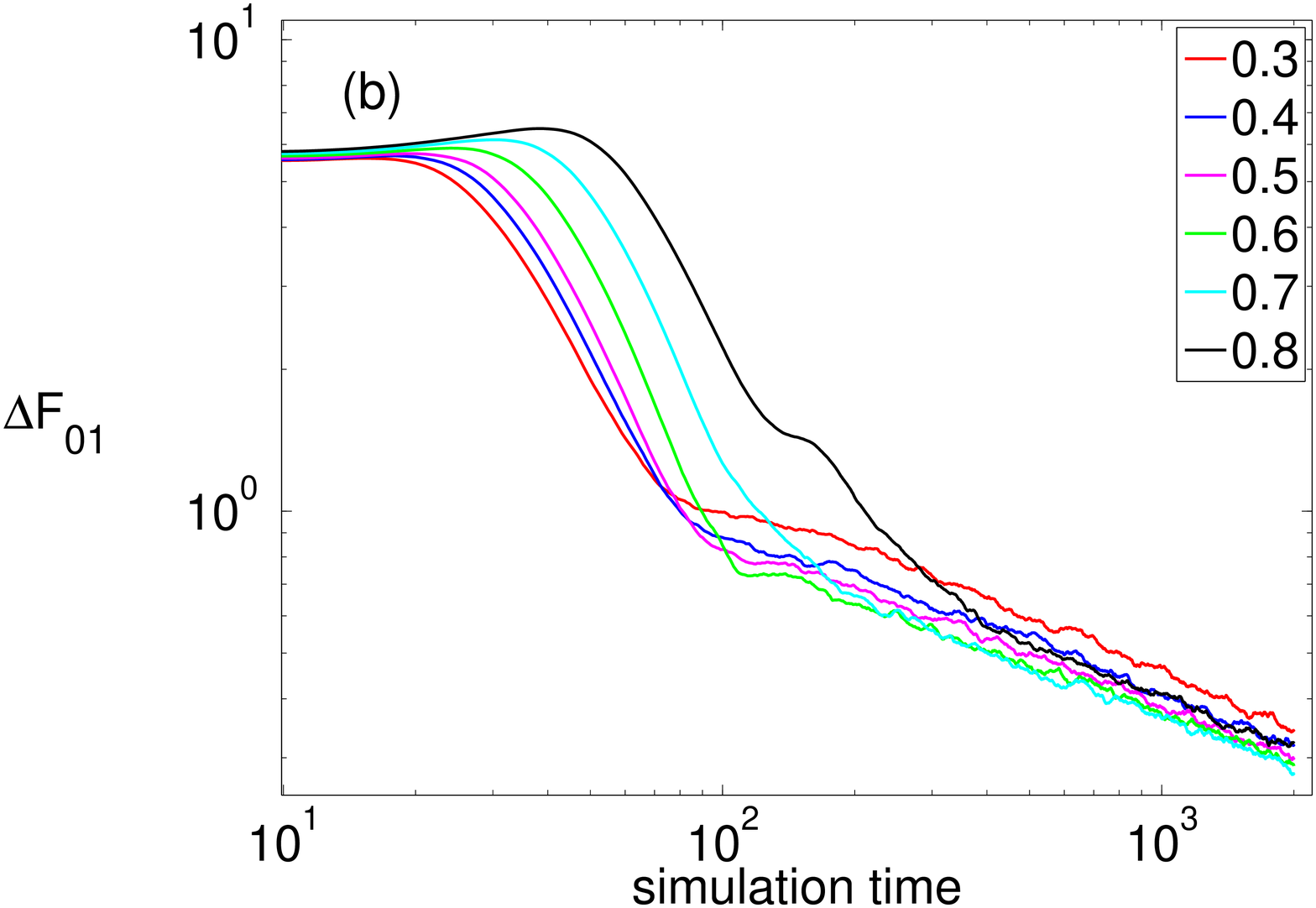}
       \end{subfigure}       
    \begin{subfigure}{0.45\textwidth}
        \includegraphics[height=1.85in]{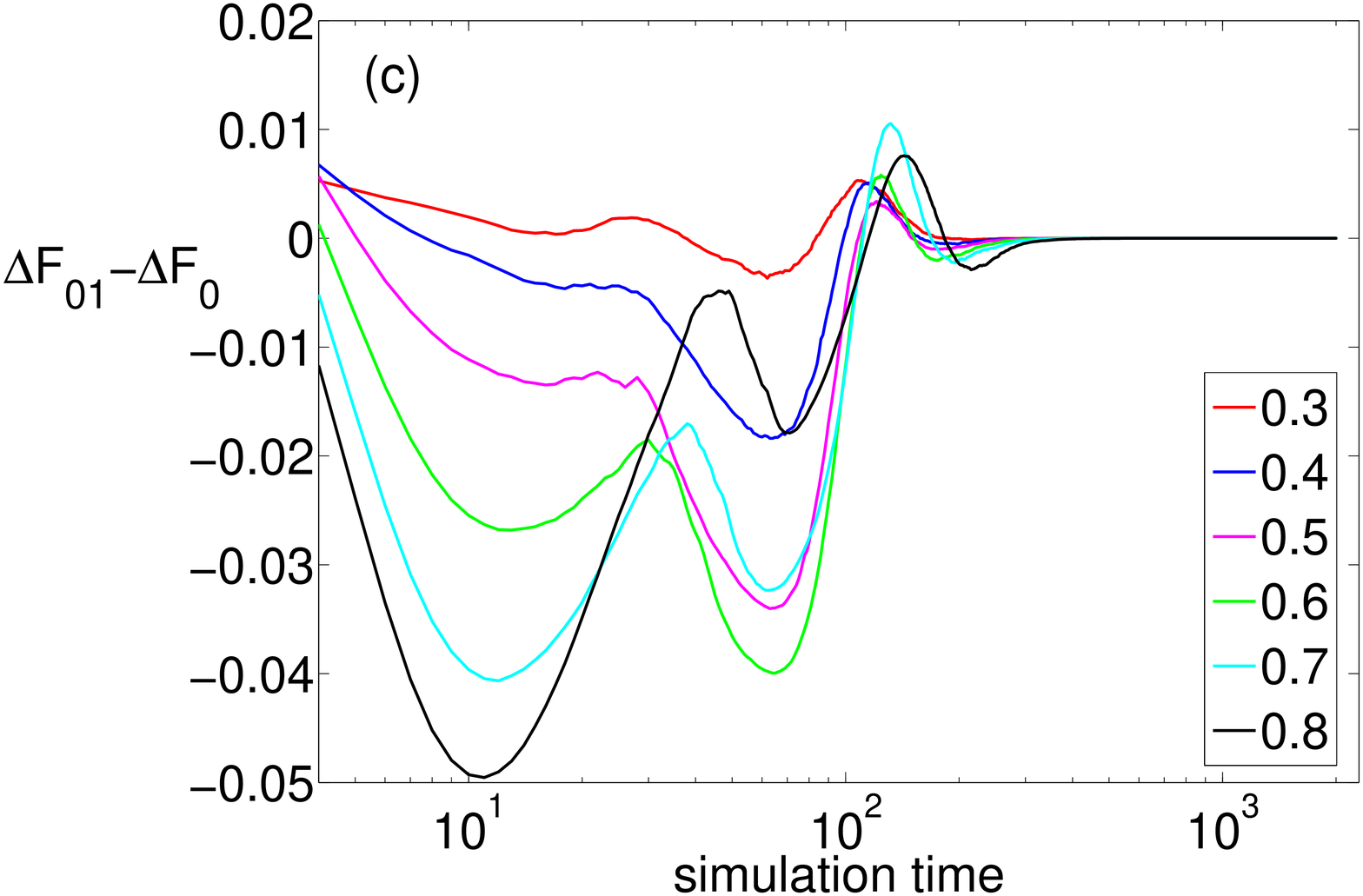}
       \end{subfigure}
\caption {(a) 1-d double well potential with a single high barrier. (b) Error in the estimated free energy using full Eq. \ref{fes} and the reference value, as a function of simulation time. Error was calculated using the metric defined in Ref. \onlinecite{adaptive} with a cutoff of 20 $k_B T$ from the global minimum.  The legend provides the respective values of the Gaussian width used to perform metadynamics simulations. (c) $\Delta F_{01} - \Delta F_{0}$, i.e. the difference in errors  with respect to the reference, as defined in (b), using the full Eq. \ref{fes} to calculate the error, versus using only the 0-th order term. All energies are in $k_B T$ units.}
\label{fes_noroll}
\end{figure}

	Recall here that  $y(s,\tau) \equiv p(s) e^{-\gamma V(s,\tau) / \Delta T}$. Using the above equation we can thus write the first-order free energy estimator for metadynamics that is valid for finite width biasing kernels to linear correction:
	\begin{equation} 
	\begin{split}
	\label{fes}
	F(s) & \equiv e^{-\beta p(s)} \equiv  F_0(s)  + F_1(s)   \\ 
	 &= - \frac{T \gamma}{\Delta T} \Big \{ V(s, \tau_0) \\
	 &+   \int ds' (G(s, s') - \delta(s, s')) V(s', \tau_0) \exp[ -\frac{\gamma V(s',\tau_0)}{\Delta T}]  \Big \}
	\end{split}
	\end{equation}
	where the subscripts denote the order of correction used to evaluate the respective quantities. Eq. \ref{fes} is one of the central results of this paper. It provides an explicit correction to metadynamics free energy for finite-width non $\delta-$like  biasing kernel, expressed as an easy to calculate convolution integral of the deposited bias. Note that the correction scales as $V(s,\tau)\exp[ -\frac{\gamma V(s,\tau)}{\Delta T}] $, and in the zeroth order, $ V(s,\tau)$ grows logarithmically with simulation time \cite{wtm,tiwary_rewt}. Thus, at least on scales comparable to the hill size or larger, any finite size-effects in fact vanish polynomially fast in simulation time. As can be seen by putting Eq. \ref{fes}  in Eq. \ref{gen_sol}, the second and higher order correction terms vanish even more quickly than the first order term, since each increasing order brings out an extra $1 \over \tau^2$ term up-front. In Fig. 1 we provide results for a 1-d asymmetric double well (Fig. 1(a)) with a single high barrier, demonstrating the precise and increasingly vanishing contribution of the second term in the Eq. \ref{fes} to the free energy evaluated from a metadynamics simulation with a variety of hill widths. In Fig. 1(b), we provide the error in the estimated free energy using the full Eq. \ref{fes} versus the reference value, as a function of simulation time.
Error was calculated using the metric defined in Ref. \onlinecite{adaptive} with a cutoff of 20 $k_B T$ from the global minimum, and averaged over 400 independent runs for each case. Irrespective of the value of the Gaussian width, there is a generic rate at which the error decays. Fig. 1(c) gives the difference in errors with respect to the reference, as defined in (b), using the full  Eq. \ref{fes}, diminished by the error using only the first term of Eq. \ref{fes}. Note that after a transient, the full estimator including first order corrections generally does give slightly less error than only the zeroth-order term - however this difference is only a negligibly small fraction of $k_B T$, and furthermore it vanishes as the simulation progresses. The relative insensitivity to the choice of Gaussian width $\sigma$ is to be underlined, even with $\sigma$ values close to the potential minimum width.

	The metadynamics simulations were performed on the potential of Fig. 1(a), with $G(s, s')  = {1 \over \sqrt{2 \pi \sigma^2} } e^{-(s-s')^2 \over 2 \sigma^2}$, and differing values of the Gaussian width $\sigma$ as marked in the legends to Figs. 1(b-c).  The biasing factor $\gamma$ equaled 15, $\omega = 5 \text{x}10^{-4} {k_B T  (\text{MD steps})^{-1} }$  and the temperature was maintained using a Langevin thermostat\cite{bussi2007accurate}. The simulation time is displayed in units of Gaussians deposited, with a Gaussian deposited every 2000 MD steps.

	\section{Conclusions}
	In this short communication, we revisited the ordinary differential equation that governs the evolution of metadynamics, and expressed it in a simple Riccati-like form valid for any generic biasing kernel. We then solved this Riccati equation using a perturbative scheme. Our solution clearly demonstrates that the perturbative effect of using finite-width hills is a negligible source of error for the method compared to the effects of discrete sampling, and therefore using the free energy estimator for metadynamics derived assuming delta-like Gaussians is well-motivated. Assuming delta-like Gaussians contributes essentially no error compared to the assumption of quasi-equilibrium sampling even when the simulation has not reached the long time limit as needed in Ref. \onlinecite{meta_ode} and for non-delta like biasing kernels. To conclude, our work further reinforces confidence in the usability of the popular metadynamics protocol for generic biasing kernels and provides a perturbative framework to systematically evaluate finite-time, finite-width corrections in the perfect sampling limit.

	\begin{acknowledgments}	
	The authors thank Ben Leimkuhler for helpful discussions.
	M.P.\ acknowledges funding from the
	National Center for Computational Design and Discovery of
	Novel Materials MARVEL and the European Union Grant
	No.\ ERC-2009-AdG-247075. 
	\end{acknowledgments}

	\bibliography{metasolution}
	\end{document}